\documentclass[aps,physrev,twocolumn,10pt,superscriptaddress,longbibliography,floatfix]{revtex4-2}
\usepackage{amsmath,amssymb,mathtools,bm,booktabs,graphicx,microtype}
\graphicspath{{./}}
\usepackage[T1]{fontenc}
\usepackage{orcidlink}
\newcommand{\erg}{\mathcal W}

\newcommand{\sym}{\mathrm{sym}}

\newcommand{\Var}{\operatorname{Var}}
\newcommand{\Cov}{\operatorname{Cov}}
\newcommand{\diag}{\operatorname{diag}}
\newcommand{\Diss}{\mathcal D}
\newcommand{\ket}[1]{|#1\rangle}
\newcommand{\bra}[1]{\langle#1|}
\begin{document}
	\title{Spectral Engineering of Charge Localization and Work Reliability in Collective Quantum Batteries}
	\author{He-Guang Xu}
	\affiliation{School of Physics and Optoelectronic Engineering, Guangdong Polytechnic Normal University, Guangzhou 510665, China}
	
	\author{G. D. de Moraes Neto~\orcidlink{0000-0003-4273-8380}}
	\email{gentil.neto@hibiuh.edu.cn}
	\affiliation{Department of Fundamental Sciences, Hainan Bielefeld University of Applied Sciences, Danzhou, Hainan 578101, China}
	\begin{abstract}
		The performance of a quantum battery depends not only on how much energy it stores, but also on how rapidly useful charge can be restored after work extraction. We show how engineering the reservoir spectrum connects these properties in an interacting collective-spin battery. Interactions separate the Dicke transition frequencies, allowing frequency-selective pumping and loss to control where excitation accumulates and how populations move along the ladder. Analytical results relate the pump-to-loss ratio to charge localization, fluctuations, and extractable work. Two reservoirs producing the same mean charge can nevertheless exhibit markedly different localization and charging times. Under repeated extraction, finite-time recharging also generates correlations between successive work outputs. At matched mean work, these correlations can reverse the ordering of single-cycle and long-time fluctuations: a protocol with larger fluctuations in individual strokes can deliver more reliable accumulated work. An exact counting relation connects these statistics to net charging-energy transfer, while relaxation modes distinguish recovery of the stationary state from recovery of useful work. Finally, local dephasing redistributes population among collective-spin sectors, leading to an exact stationary symmetry-dilution law and a finite operating window for work extraction under a collective transverse controller.
	\end{abstract}
	\maketitle
	
	\section{Introduction}\label{sec:intro}
	
	Quantum batteries exploit controllable quantum systems to store energy that can subsequently be recovered as useful work\cite{Khadim2026}. Their performance depends on the energy accumulated during charging, the work accessible from the prepared state, and the time required to complete the process. Coherent charging protocols use external fields or auxiliary quantum systems to transfer energy through unitary dynamics, with collective interactions offering enhanced charging rates and opportunities to control the resulting many-body states~\cite{AlickiFannes2013,Campaioli2024RMP,Ferraro2026Perspective}. The charged state can then be subjected to a controlled operation that transfers energy to an external system. Repeated operation introduces an additional consideration: the battery must recover its useful charge after each extraction, making the dynamics of energy replenishment central to its performance.
	
	Dissipative charging provides another route to preparing and maintaining energetic quantum states. Here, coupling to an externally powered environment supplies energy while relaxation channels continuously redistribute or remove excitations. Engineering these competing processes can stabilize nonequilibrium states with ergotropy---work extractable by a cyclic unitary transformation---and sustain charging in the presence of losses~\cite{Barra2019,QuachMunro2020,Carrasco2022,ZhangJing2025StableCharging,Yang2026ErgotropyActivation}. Measurement-feedback charging in waveguide-QED systems provides a distinct route to controlling stored energy and ergotropy~\cite{Yin2026}; the present model uses the frequency dependence of the reservoir response to select the occupied regions of the Dicke ladder. The stationary occupation distribution is determined by the balance of pumping and loss, whereas the transition rates also govern the time required to approach that distribution. Reservoir engineering consequently provides control over both the properties of the stored charge and its recovery following work extraction.
	
	Collective-spin systems offer a particularly useful realization of this approach. An ensemble of identical two-level constituents supports a symmetric Dicke ladder whose neighboring transitions exhibit occupation-dependent collective enhancement~\cite{Dicke1954}. Interactions between the constituents can separate the transition frequencies while preserving the collective spin-flip matrix elements, allowing a frequency-dependent reservoir to address different regions of the ladder. This mechanism underlies quantum-engineered superabsorption and related approaches to selective collective excitation~\cite{Higgins2014}. Interacting Dicke batteries and dissipative collective charging have also been investigated as routes to enhanced energy storage and charging power~\cite{Dou2022InteractingDicke,Pokhrel2025,Dias2026,Zhou2026,QuachMunro2020}. Oscillator-assisted multi-atom charging likewise connects collective dynamics to extractable work~\cite{Huang2026AQTCollective}. In the present setting, the interaction separates neighboring Dicke transitions so that the reservoir spectrum controls the rates at which population moves toward, accumulates around, and escapes from selected excited states.
	
	The connection between stationary charge and charging dynamics becomes especially important during repeated energy extraction. A reservoir may concentrate population within a narrow range of excitations while requiring a substantial time to restore that distribution after a work stroke. The extracted energy also fluctuates from cycle to cycle, and incomplete recharging can generate temporal correlations between successive outcomes. Previous studies have established the importance of extractable-work fluctuations, charging precision, and repeated-cycle statistics in quantum batteries and quantum thermal machines~\cite{GarciaPintos2020Fluctuations,Barra2022Cycle,Bakhshinezhad2024,Rinaldi2025,Rinaldi2026,Mohan2026,Garcez2026}. For an anharmonic Dicke ladder, the spectral placement of the charging reservoir determines the transport pathways connecting successive operating states. This links the localization of stationary charge, the timescale of its recovery, and the fluctuations of accumulated work. The relaxation spectrum supplies a further control coordinate: different preparations can approach the same charged stationary distribution at different rates because they excite different relaxation modes, as in Mpemba-type dynamics~\cite{Ares2025Perspective,Medina2025}. The work recovered by a particular extraction operation also depends on which occupations these modes redistribute.
	
	The collective structure also influences the stability of the charged state under environmental noise. The symmetric Dicke ladder occupies a restricted sector of the full many-spin Hilbert space, and independent local processes can transfer population into sectors with smaller total spin. Such processes affect the work accessible through collective operations even when substantial excitation energy remains in the ensemble. The interplay between local and collective dissipation has been investigated in quantum-optical spin ensembles and quantum batteries~\cite{Shammah2018,Canzio2025,Khoudiri2026Dark}. Symmetry-protected excitonic states provide a related example in which the accessibility and lifetime of stored charge depend on dissipative selection~\cite{Khodadad2025AQTDark}. In the Dicke battery, local channels redistribute excitation among total-spin sectors and thereby change the work available to the collective controller. The same physical sequence can therefore be followed from the spectrum that selects a charged state, through the transition rates that restore it, to the correlations and symmetry sectors sampled by the work controller.
	
	For a frequency-selectively pumped Dicke battery, the stationary occupation ratios determine where charge accumulates, while the individual transition rates determine how readily it can be restored. We examine the consequences for ergotropy and target-access time, then for the correlations between work outputs when extraction and recharging are repeated. Two reservoirs with the same mean charge can differ in localization and charging speed; at matched mean extracted work, their single-stroke and accumulated-work fluctuations can even have opposite orderings. Local dephasing adds a distinct constraint by moving population between total-spin sectors and changing its accessibility to a collective work controller. These connections place reservoir-selected charging alongside coherent and dissipative battery protocols~\cite{Khadim2026,Rinaldi2026}.
	
	\begin{figure}[t]
		\centering
		\includegraphics[width=\columnwidth]{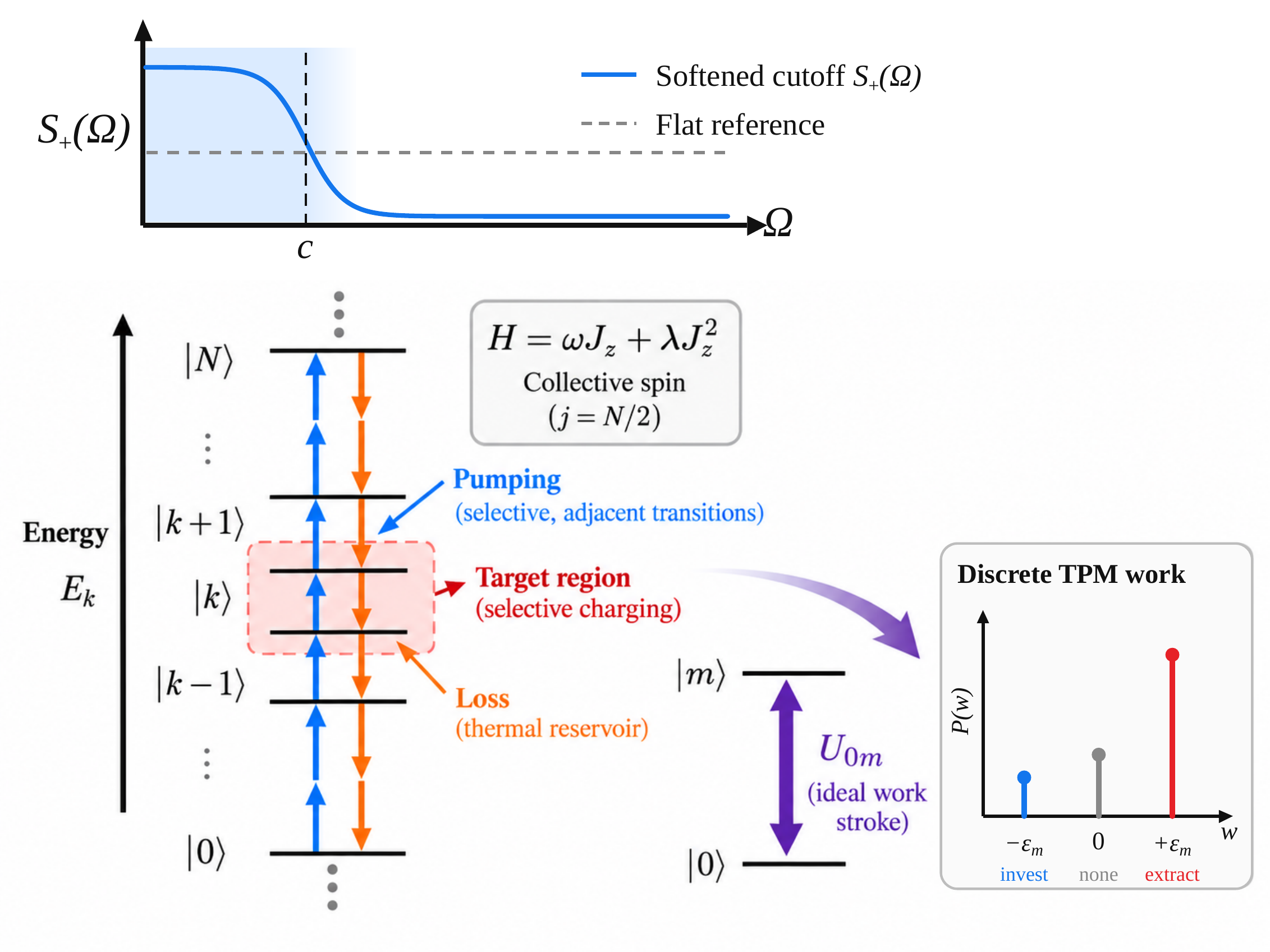}
		\caption{Spectrally selective charging and work extraction in an interacting Dicke battery. The schematic softened-cutoff pump $S_+(\Omega)$ has threshold center $c$ and a nonzero high-frequency floor; the dashed flat spectrum is a shape reference, not a rate-matched comparison [Eq.~\eqref{eq:cutoff} and Appendix~\ref{app:numerical-reservoirs}]. Interaction-resolved Dicke edges experience upward pumping (blue) and loss (orange), which determine occupation near a target region. Following charging, the ideal unitary $U_{0m}$ exchanges $\ket0$ and $\ket m$. With extracted work positive, the \emph{discrete} TPM outcomes $w=-\epsilon_m,0,+\epsilon_m$ have probabilities $p_0^-$, $1-p_0^--p_m^-$, and $p_m^-$; stem heights are illustrative.}
		\label{fig:mechanism}
	\end{figure}
	
	\section{Collective battery and operating protocol}\label{sec:model}
	
	\subsection{Collective battery and operating Hamiltonian}
	
	We consider a battery formed by $N$ identical two-level constituents coupled to a powered pumping reservoir, an independent loss reservoir, and an external work controller. Charging transfers energy from the powered reservoir to the collective spin, while a controlled unitary operation extracts energy after each charging contact. The couplings can be switched off to store the charge in isolation. With $\hbar=1$, the Hamiltonian organizing these operations is
	\begin{align}\label{eq:Htot}
		H_{\rm tot}(t)={}&H_B^{(N)}+H_{\rm p}+H_{\ell}\nonumber\\
		&+g_{\rm c}(t)\bigl[V_{\rm p}+V_{\ell}+H_{\rm src}(t)\bigr]\nonumber\\
		&+g_{\rm w}(t)H_{\rm w}(t).
	\end{align}
	Here $H_B^{(N)}$ is the battery Hamiltonian, $H_{\rm p}$ and $H_{\ell}$ describe the two reservoirs, and $V_{\rm p}$ and $V_{\ell}$ couple them to the battery. The source term $H_{\rm src}(t)$ supplies energy to the pumping reservoir. During charging, the switching functions take $g_{\rm c}=1$ and $g_{\rm w}=0$; during extraction they take $g_{\rm c}=0$ and $g_{\rm w}=1$. Setting both to zero isolates the battery. The repeated protocol alternates charging and extraction, with an optional storage interval between them.
	
	Collective absorption and emission are generated by $J_\pm=\sum_{i=1}^N\sigma_\pm^{(i)}$, with $J_z=\frac12\sum_{i=1}^N\sigma_z^{(i)}$. For example, each reservoir coupling can be represented as $V_\nu=J_+\otimes B_\nu+J_-\otimes B_\nu^\dagger$ for $\nu\in\{\mathrm p,\ell\}$. The pump-source dynamics and reservoir correlations determine the frequency-dependent upward and downward transition strengths used below. For the collective charging model we begin in the fully symmetric sector $J=N/2$, whose Dicke states $\ket{k}$ carry $k=0,\ldots,N$ excitations~\cite{Dicke1954,Higgins2014}. Their collective matrix elements are
	\begin{align}
		J_z\ket{k}&=\left(k-\frac N2\right)\ket{k},\nonumber\\
		J_-\ket{k}&=\sqrt{k(N-k+1)}\,\ket{k-1}.
	\end{align}
	
	The symmetric manifold contains $N+1$ levels, and its diagonal dynamics admit an $N+1$-state rate description. Single multilevel batteries provide closely related settings for dissipative stabilization and controlled charging~\cite{ZhangJing2025StableCharging,Huang2026Multilevel,Gemme2024}. The full spin ensemble also contains sectors with smaller total spin. Local processes connect these sectors, and their populations respond differently to a collective work controller~\cite{Shammah2018,Canzio2025,Khoudiri2026Dark}, as examined in Sec.~\ref{sec:leakage}.
	
	The battery Hamiltonian contains the constituent excitation energy and an all-to-all longitudinal interaction,
	\begin{equation}\label{eq:H}
		H_B^{(N)}=\omega J_z+\lambda_NJ_z^2.
	\end{equation}
	The quadratic term has the one-axis-twisting form familiar from collective-spin dynamics~\cite{KitagawaUeda1993,Dou2022InteractingDicke}. Its microscopic spin content follows from
	\begin{equation}
		J_z^2=\frac N4+\frac12\sum_{i<j}\sigma_z^{(i)}\sigma_z^{(j)},
	\end{equation}
	so $\lambda_N$ sets the strength of identical longitudinal pair interactions. Such interactions shift the frequencies addressed by a reservoir while leaving the Dicke transition matrix elements unchanged, a basis for spectrally selective collective absorption~\cite{Higgins2014}. Transverse interactions provide a complementary charging mechanism by generating spin squeezing and nonlinear collective torques~\cite{Yan2026SpinSqueezing}; here the longitudinal interaction makes neighboring ladder transitions separately addressable by the reservoirs. We consider two ways of varying the system size: $\lambda_N=\chi/N$ keeps the interaction energy extensive, as in interacting collective-battery models~\cite{Dou2022InteractingDicke}; $\lambda_N=\lambda_0$ holds the coupling of each pair fixed, as in finite-spin one-axis-twisting dynamics~\cite{KitagawaUeda1993}. Their different large-system behavior is examined in Sec.~\ref{sec:stationary}, with the finite-size parameter choices specified in Appendix~\ref{app:numerical-reservoirs}. Related multilevel and multi-photon charging mechanisms are discussed in~\cite{Huang2026Multilevel,Gemme2024,Zahia2025}.
	
	After subtracting the energy of $\ket0$, the Dicke energies, transition frequencies, and collective strengths are
	\begin{align}\label{eq:spec}
		\epsilon_k&=\omega k+\lambda_N k(k-N),\\
		\Omega_k&=\epsilon_k-\epsilon_{k-1}
		=\omega+\lambda_N(2k-N-1),\nonumber\\
		C_k&=|\bra{k}J_+\ket{k-1}|^2=k(N-k+1).\nonumber
	\end{align}
	The interaction separates neighboring transition frequencies by $\Omega_{k+1}-\Omega_k=2\lambda_N$, while $C_k$ retains its collective enhancement near the middle of the ladder. We work in the ascending-energy regime $\Omega_k>0$, equivalent to $\omega>|\lambda_N|(N-1)$, so each upward transition increases the stored energy.
	
	\subsection{Spectrally selective charging}
	
	Powered pumping and collective dissipation provide established approaches to charging open multiqubit batteries~\cite{Dias2026,Zhou2026}. Here the powered reservoir induces upward Dicke transitions and the loss reservoir induces downward transitions, each resolved by the corresponding transition frequency. Let $S_+(\Omega)$ and $S_-(\Omega)$ denote their effective rate spectra. When neighboring Bohr frequencies are resolved on the dissipative scale, the individual edges follow the frequency-resolved weak-coupling description~\cite{Davies1974,Trushechkin2021}. For a state diagonal in the Dicke basis, with $p_k=\bra{k}\rho\ket{k}$, the populations evolve as
	\begin{align}\label{eq:rates}
		u_{k-1}&=C_kS_+(\Omega_k),\qquad
		d_k=C_kS_-(\Omega_k),\\
		\dot p_k&=u_{k-1}p_{k-1}+d_{k+1}p_{k+1}
		-(u_k+d_k)p_k,\nonumber
	\end{align}
	where rates beyond the ladder endpoints vanish. We denote the population generator by $L$ and the finite charging-contact propagator by $C_\tau=e^{L\tau}$. For the numerical comparisons we impose $\Gamma_{\max}/\Delta\Omega\leq0.04$, where $\Gamma_{\max}=\max_k(u_k+d_k)$ and $\Delta\Omega=2|\lambda_N|$ is the neighboring-transition spacing. On this scale, the dissipative linewidths remain well separated from the resolved Dicke frequencies. The corresponding quantum master equation is given in Appendix~\ref{app:generator}. 
	
	For positive edge rates, the stationary populations $\bm\pi$ obey local balance,
	\begin{equation}\label{eq:pi}
		\frac{\pi_k}{\pi_{k-1}}=\frac{u_{k-1}}{d_k}
		=r(\Omega_k),\qquad
		r(\Omega)=\frac{S_+(\Omega)}{S_-(\Omega)},
	\end{equation}
	so $\pi_k=Z^{-1}\prod_{j=1}^k r(\Omega_j)$, with $Z$ fixed by normalization. Spectral ratios therefore control which occupations accumulate, while the collective factors $C_k$ set the scale of the corresponding transport rates. In a passive thermal contact, spectral shaping can redistribute edge currents while thermal detailed balance retains the Gibbs stationary state, as in the collective Otto engine of Ref.~\cite{DamasNeto2026ThermodynamicRouting}. In the powered battery, the frequency dependence of $S_+(\Omega)/S_-(\Omega)$ also selects the stationary charge itself. At stationarity, $u_{k-1}\pi_{k-1}=d_k\pi_k$: opposite population fluxes balance on each edge although absorption and emission continue.
	
	The pump and loss carry gross edge-energy currents $j_k^+=\Omega_ku_{k-1}p_{k-1}$ and $j_k^-=\Omega_kd_kp_k$, giving the battery-energy balance
	\begin{equation}\label{eq:power}
		\frac{d}{dt}\langle H_B\rangle
		=\sum_{k=1}^N\bigl(j_k^+-j_k^-\bigr).
	\end{equation}
	At stationarity, the equal and opposite edge currents sustain a nonequilibrium energy throughflow from the powered pump to the loss reservoir while the stored battery energy remains constant. The gross energies transferred over a charging contact are
	\begin{equation}\label{eq:gross_charge_energies}
		Q_\pm(\tau)=\int_0^\tau dt\sum_{k=1}^N j_k^\pm(t),\qquad
		\Delta E_B=Q_+(\tau)-Q_-(\tau).
	\end{equation}
	The gross currents measure energy transferred through the battery. The rate-capacity constraint compares the available transition rates, whereas the free energy needed to maintain an active pump depends on the physical source that produces $S_+(\Omega)$. During isolated storage the battery evolves under $H_B^{(N)}$; its energy-diagonal populations, and hence its stored energy, remain constant.
	
	We compare three upward pump shapes against the same loss spectrum: a flat response, a Lorentzian peak, and a softened cutoff. With center frequency $c$, width $\zeta>0$, and positive off-resonant floor $b$, they are
	\begin{align}\label{eq:cutoff}
		f_{\rm cut}(\Omega;c,\zeta)&=b+\frac{1}{1+\exp[(\Omega-c)/\zeta]},\\
		f_{\rm pole}(\Omega;c,\zeta)&=b+\frac{\zeta^2}{(\Omega-c)^2+\zeta^2},\nonumber\\
		f_{\rm flat}(\Omega)&=1.\nonumber
	\end{align}
	The softened cutoff supplies strong pumping at lower frequencies and a weaker background above the threshold. Its center $c$ determines which Dicke transitions lie on either side of the crossover, while $\zeta$ sets how sharply they are distinguished and $b$ retains residual excitation at high frequency. The battery samples this effective source spectrum at its Bohr frequencies. Structured cavity environments provide another route to modifying charge and extractable work through reservoir dynamics~\cite{Hadipour2024AQTStructured}; frequency-selective passive contacts have also been analyzed for collective Otto cycles~\cite{DamasNeto2026ThermodynamicRouting}. In the present battery the powered source instead supplies the upward transitions.
	
	For the equal-capacity comparisons, the pump amplitude fixes the total upward edge rate $B_N=\sum_{k=1}^N u_{k-1}$. The comparison therefore tests how spectral placement redistributes a given kinetic capacity among the transitions, a distinction also relevant to control-constrained charging architectures~\cite{Siddiqui2026Resource}. Scaling all upward and downward rates by a common factor sets the resolved-transition timescale without changing the stationary ratios. Appendix~\ref{app:numerical-reservoirs} specifies the rates and parameters. Figures~\ref{fig:stationary_map}--\ref{fig:work_map} use the softened cutoff with controls $c/\omega$ and $\zeta/\omega$; the flat and Lorentzian spectra provide additional references.
	
	In Fig.~\ref{fig:mechanism}, the positive interaction shifts successive Dicke edges to higher frequency. The cutoff consequently weakens pumping above a selected part of the ladder while loss remains active, creating an occupation peak where the local pump-to-loss ratio crosses unity. The target is populated through dissipative transport along the ladder rather than by a coherent drive that addresses a selected state directly.
	
	\subsection{Stored energy, number fluctuations, and charging times}
	
	A charged state with Dicke populations $\pi_k$ stores average energy $E_B=\sum_k\epsilon_k\pi_k$. The maximum work available through cyclic unitaries acting within the symmetric sector is its ergotropy~\cite{Allahverdyan2004,AlickiFannes2013},
	\begin{equation}\label{eq:erg}
		\erg_\sym(\pi)=\sum_{k=0}^N\epsilon_k(\pi_k-\pi_k^\downarrow),
	\end{equation}
	where $\pi_k^\downarrow$ are the populations rearranged in decreasing order against the ascending energy levels. The work delivered by the specific control stroke below is determined separately from its actual pre-extraction state.
	
	We locate the charge through $x=\langle k\rangle/N$ and characterize occupation fluctuations using
	\begin{equation}\label{eq:rcss}
		F=\frac{\Var(k)}{\langle k\rangle},\qquad
		R_{\mathrm{CSS}}=\frac{\Var(k)}{Nx(1-x)}.
	\end{equation}
	The denominator of $R_{\mathrm{CSS}}$ is the binomial variance of a spin-coherent state with the same mean occupation. Thus $R_{\mathrm{CSS}}<1$ identifies sub-binomial Dicke-number fluctuations, whereas $R_{\mathrm{CSS}}>1$ identifies a broader occupation distribution. This comparison remains meaningful near saturation, where the Fano factor can be small even for a broad distribution relative to its binomial reference. Transverse spin squeezing probes different correlations.
	
	We characterize charging dynamics by the first-passage time $T_m$ for a trajectory starting from $\ket0$ to reach a target level $m$, and by the stationary weight $\pi_{\mathrm{win}}=\sum_{|k-m|\leq\delta}\pi_k$ in a target window of integer half-width $\delta\geq0$. The contact duration $\tau$ fixes when extraction occurs; $\mathbb E_0T_m$ measures target accessibility and $\pi_{\mathrm{win}}$ characterizes the prepared charge localization.
	
	\subsection{Work extraction and repeated operation}
	
	After a charging contact we apply a controlled unitary to the battery. Energy-basis permutations provide a well-defined class of finite-dimensional extraction operations~\cite{FrancicaDellAnna2024Risk}. Our reference stroke exchanges $\ket0$ and $\ket m$ and preserves the occupations of the remaining Dicke levels. In an interaction picture resonant with their energy difference $\epsilon_m$, it is generated by
	\begin{equation}
		H_{{\rm w},I}(t)=g_m(t)\bigl(\ket0\bra m+\ket m\bra0\bigr),
		\qquad \int g_m(t)\,dt=\frac\pi2.
	\end{equation}
	The pulse realizes the exchange up to phases and defines the work controller in Eq.~\eqref{eq:Htot}.  This exchange isolates a three-outcome work process whose repeated statistics can be traced directly to the charging propagator. The induced population permutation is $D_{jk}=\delta_{j,f(k)}$, with $f(0)=m$, $f(m)=0$, and $f(k)=k$ for the remaining levels. Section~\ref{sec:leakage} also considers a collective $J_x$ controller that acts across total-spin sectors.
	
	Two-point-measurement (TPM) work resolves the energy decrease of the battery under this specified stroke. With positive work denoting extraction,
	\begin{equation}\label{eq:w}
		w(k)=\epsilon_k-\epsilon_{f(k)}=
		\begin{cases}
			+\epsilon_m,&k=m,\\
			-\epsilon_m,&k=0,\\
			0,&\mathrm{otherwise}.
		\end{cases}
	\end{equation}
	A pre-stroke occupation of $m$ releases energy, while occupation of zero requires the controller to invest the same energy. If $\bm p^-$ is the probability distribution immediately before extraction, the mean output is $\mu=\epsilon_m(p_m^--p_0^-)$. Figure~\ref{fig:mechanism} displays the three discrete work outcomes with weights $p_m^-$, $p_0^-$, and $1-p_m^--p_0^-$. This ideal exchange differs from the collective transverse drive studied under local noise in Sec.~\ref{sec:leakage}.
	
	The cycle begins immediately after extraction. The charging propagator and work permutation produce the cycle map $P=DC_\tau$. In the periodic regime, the post-stroke distribution satisfies $P\bm q=\bm q$, and the pre-stroke state is $\bm p^-=C_\tau\bm q$. The work-tilted map is
	\begin{equation}\label{eq:tilt}
		P_s=D\diag(e^{sw_0},\ldots,e^{sw_N})C_\tau.
	\end{equation}
	Writing $\Lambda(s)$ for its dominant eigenvalue analytic near $s=0$, with $\Lambda(0)=1$, the mean extracted work and asymptotic variance accumulated per cycle are
	\begin{align}\label{eq:fcs}
		\mu&=\left.\partial_s\ln\Lambda(s)\right|_{s=0},\\
		D_W&=\left.\partial_s^2\ln\Lambda(s)\right|_{s=0}\nonumber\\
		&=V_1+2\sum_{r=1}^\infty\Cov(w_0,w_r),\nonumber
	\end{align}
	where $V_1=\langle w^2\rangle-\mu^2$ is the variance of one work stroke in the stationary cycle. For a mixing cycle, $\Var(\sum_{j=1}^M w_j)/M\to D_W$ as $M\to\infty$. We also track the probabilities of extracting and investing work.
	
	The ratio $\mu^2/(D_W\tau)$ expresses reliability per charging-contact time. When the extraction control occupies a duration $t_{\rm w}$ and an isolated storage interval lasts $t_{\rm s}$, the full elapsed cycle time is $\tau+t_{\rm w}+t_{\rm s}$. The collective-drive comparison in Sec.~\ref{sec:leakage} retains its finite pulse duration. Repeated operation thus connects spectral charge preparation to correlations between successive work events.
	
	\section{Spectral shaping of stationary charge}\label{sec:stationary}
	
	The stationary charge follows from the competition between upward pumping and downward loss at each Dicke transition. When the edge ratio $r(\Omega_k)$ exceeds unity, occupation increases from $k-1$ to $k$; when it falls below unity, occupation decreases. An interior peak therefore forms where the relative gain changes from upward-favored to downward-favored transport. Its position depends on the crossing of $r(\Omega)$ through unity, while its width reflects how rapidly the ratio varies across neighboring transitions. An interaction-resolved reservoir can control both properties through its spectral profile~\cite{Pokhrel2025,Dias2026}.
	
\begin{figure}[t]
		\centering
		\includegraphics[width=\columnwidth]{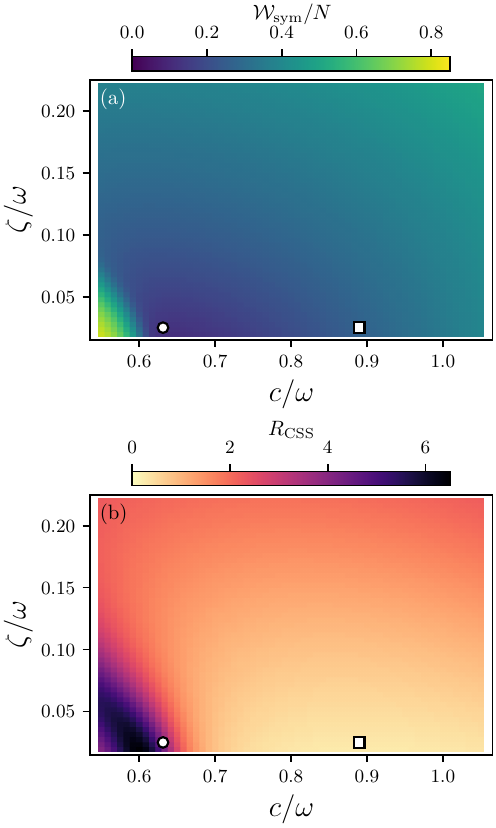}
		\caption{Stationary charge selected by the softened-cutoff reservoir at $N=20$, with $\lambda_N=\chi/N$, $\chi=0.48$, and $\omega=1$. The maps show (a) symmetric-sector ergotropy per spin $\erg_\sym/N$ and (b) excitation-number variance relative to the binomial reference, $R_{\mathrm{CSS}}=\Var(k)/[Nx(1-x)]$, where $x=\langle k\rangle/N$. The horizontal and vertical axes are the normalized cutoff center $c/\omega$ and width $\zeta/\omega$, respectively. Pump-rate capacity and the loss response are shared across the map. Circle and square identify the two profiles with equal stationary mean occupation $x\simeq0.42$; their localization and ergotropy differ. The plotted parameters satisfy the resolved-transition rate ceiling.}
		\label{fig:stationary_map}
	\end{figure}

	Figure~\ref{fig:stationary_map} shows how the reservoir spectrum controls the extractable energy and the width of the stationary charge distribution. At equal mean occupation and upward-rate capacity, the upper-center cutoff approximately doubles the ergotropy and sharply suppresses excitation-number fluctuations. The lower-center cutoff instead leaves a broad high-occupation tail. The difference follows from the local decrease of the pump-to-loss ratio across the occupied transitions: a steeper spectral balance confines charge more tightly around the target.
	
	A flat upward spectrum provides a separate reference. Its stationary population still depends on frequency through the structured loss response, but retuning its amplitude to match the mean charge produces a broad distribution, with $R_{\mathrm{CSS}}\simeq7.6$ and very little sector ergotropy. This amplitude-matched reference uses a different upward-rate budget; the two cutoff reservoirs in Fig.~\ref{fig:stationary_map} share the same budget. Changing the contact duration alters the approach to stationarity while preserving the stationary state.
	
	Sampled single-Lorentzian pumps reveal a complementary charging regime: their largest sector ergotropy is achieved near full occupation rather than around an interior excitation. Near saturation, the binomial reference variance becomes small, so the normalized variance can remain appreciable even for a highly charged battery. The cutoff reservoir instead stabilizes a sharply localized state at intermediate occupation. These protocols thus select different regions of the charging landscape.
	
	\subsection{Low-excitation and finite-density limits}
	
	The behavior at large system size depends on the occupied portion of the Dicke ladder. Writing $n=a^\dagger a$, the exact Holstein--Primakoff representation in the truncated symmetric sector is $J_z=n-N/2$, $J_+=a^\dagger\sqrt{N-n}$, and $J_-=\sqrt{N-n}\,a$. After shifting the ground-state energy,
	\begin{equation}\label{eq:hp}
		H_B-E_0=(\omega-N\lambda_N)n+\lambda_N n^2,
		\qquad 0\leq n\leq N.
	\end{equation}
	For extensive coupling $\lambda_N=\chi/N$ and a fixed excitation number $n=O(1)$, the limiting Hamiltonian is harmonic, with frequency $\omega-\chi$. Its stable, phase-insensitive linear gain and loss dynamics yield a geometric stationary distribution with $\Var(n)=\bar n(1+\bar n)$ and a passive state. This harmonic limit provides the reference for the finite-density regime, where interactions continue to resolve distinct charging transitions.
	
	A battery holding a finite fraction of its capacity follows a different limit. With $k=Nx$, $0<x<1$, and $\lambda_N=\chi/N$, both the excitation number and the collective matrix elements remain extensive in their natural scales. The energy per spin and the transition frequency approach
	\begin{equation}\label{eq:macro}
		\frac{\epsilon_{Nx}}{N}\longrightarrow e(x)=\omega x+\chi(x^2-x),
		\qquad \Omega(x)=\omega+\chi(2x-1).
	\end{equation}
	Thus a stationary state concentrated near an interior fraction $x$ continues to sample occupation-dependent transition frequencies as $N$ increases.
	
	The reference pump normalization in Appendix~\ref{app:numerical-reservoirs} also has a well-defined continuum limit. For a fixed smooth positive spectral shape $f$,
	\begin{align}\label{eq:limit_ratio}
		A_N&=\frac{B_{N,0}}{\sum_{k=1}^N C_k f(\Omega_k)}
		\longrightarrow A_\infty,\nonumber\\
		A_\infty&=\frac{0.065/6}{\int_0^1dy\,y(1-y)f[\Omega(y)]},\\
		r_\infty(\Omega)&=\frac{A_\infty f(\Omega)}{S_{-,0}(\Omega)}.\nonumber
	\end{align}
	Here $\sum_{k=1}^N C_k=N(N+1)(N+2)/6$, and the common weak-rate factor cancels from the stationary ratio at every finite $N$. The resulting smooth positive $r_\infty$ defines the macroscopic stationary distribution through
	\begin{equation}\label{eq:ldp}
		\pi_{Nx}\asymp\exp[N\Phi(x)],\qquad
		\Phi(x)=\int_0^x\ln r_\infty[\Omega(y)]\,dy,
	\end{equation}
	with normalization and subexponential corrections understood. A unique, stable interior maximum $x_*$ has
	$r_\infty[\Omega(x_*)]=1$ and $2\chi\,\partial_\Omega\ln r_\infty(\Omega_*)<0$. Expansion about that maximum gives
	\begin{align}\label{eq:variance_macro}
		\lim_{N\to\infty}\frac{\Var(k)}{N}
		&=-\frac{1}{2\chi\partial_\Omega\ln r_\infty(\Omega_*)},\\
		F_\infty&=-\frac{1}{2\chi x_*\partial_\Omega\ln r_\infty(\Omega_*)}.\nonumber
	\end{align}
	When the ascending ladder has a single such maximum and the stationary width is $O(\sqrt N)$, passive rearrangement places the largest eigenvalues among the lowest $O(\sqrt N)$ energy levels. The passive energy per spin vanishes in this limit, leaving $\erg_\sym/N\to e(x_*)$. The local slope of the spectral gain-to-loss ratio therefore controls the number fluctuations of a macroscopically charged state, while the peak position determines its leading stored ergotropy.
	
	An exponential ratio makes these relations transparent. Taking $r(\Omega)=\exp[-a(\Omega-\omega)]$ with $a>0$ and $\chi>0$ yields
	\begin{align}\label{eq:gauss}
		\pi_k&\propto\exp\!\left[-\frac{a\chi}{N}
		\left(k-\frac N2\right)^2\right],\\
		F_\infty&=\frac1{a\chi},\qquad
		R_{\mathrm{CSS},\infty}=\frac2{a\chi}.\nonumber
	\end{align}
	The finite-size results converge toward the predicted fluctuation width and extensive ergotropy (Appendix Fig.~\ref{fig:app_largeN}). This soluble example isolates the role of the spectral slope; the pump maps in Fig.~\ref{fig:stationary_map} use the budget-normalized softened cutoff.
	
	The two interaction continuations also have different spectral-resolution properties. With the extensive coupling, neighboring transition frequencies approach one another as $2\chi/N$; under the specified resolved-edge rate ceiling, target-access time consequently grows at least quadratically with $N$ (Sec.~\ref{sec:kinetics}). At fixed pair coupling $\lambda_N=\lambda_0$, the transition spacing stays constant over the allowed ascending-ladder size window. A size-dependent choice $\omega_N=N\lambda_0+\omega_*$ instead produces the Kerr-oscillator Hamiltonian $\omega_*n+\lambda_0 n^2$ at low excitation. A localized stationary charge can therefore retain a finite-density limit even as the spectral resolution needed to prepare it constrains the accessible charging times.
	
	\section{Charging kinetics: spectral access and recharge}\label{sec:kinetics}
	
	The stationary distributions of Sec.~\ref{sec:stationary} characterize where excitation accumulates after a long reservoir contact. Reaching a chosen Dicke occupation from an initially uncharged battery probes another aspect of the same transport. Every upward step carries population across one edge of the ladder, while downward transitions produce excursions that must be retraced before the target is reached. Pump spectral placement therefore affects the time needed to access a charged state as well as the distribution maintained around it~\cite{Beder2026}.
	
	For an initial occupation $k=0$, let $T_m$ be the first time the trajectory visits a target level $m$. The nearest-neighbor generator admits an exact backward recursion. With $\delta_i=\mathbb E_iT_{i+1}$,
	\begin{equation}\label{eq:mfpt}
		\delta_0=\frac{1}{u_0},\qquad
		\delta_i=\frac{1+d_i\delta_{i-1}}{u_i},\qquad
		\mathbb E_0 T_m=\sum_{i=0}^{m-1}\delta_i.
	\end{equation}
	Using the stationary edge balance $u_i\pi_i=d_{i+1}\pi_{i+1}$, the same mean first-passage time can be written as
	\begin{equation}\label{eq:mfpt_traffic}
		\mathbb E_0T_m=\sum_{i=0}^{m-1}
		\frac{\sum_{j=0}^{i}\pi_j}{u_i\pi_i}.
	\end{equation}
	The numerator counts the stationary probability lying below each edge, while $u_i\pi_i$ is the upward probability flux across it. The time to reach $m$ therefore depends on the traffic through every preceding edge, even when two reservoirs yield similar occupation near the target.
	
	The resolved-transition condition also sets a scale for target access. A nearest-neighbor path from $0$ to $m$ contains at least $m$ jumps. Writing $\Gamma_{\max}=\max_k(u_k+d_k)$ for the largest escape rate, with endpoint rates understood to vanish, gives
	\begin{equation}\label{eq:mfptbound}
		\mathbb E_0T_m\geq\frac{m}{\Gamma_{\max}}.
	\end{equation}
	For the rate ceiling $\Gamma_{\max}\leq\alpha\Delta\Omega=2\alpha|\lambda_N|$ imposed on the resolved-edge model, the extensive interaction $\lambda_N=\chi/N$ yields $\mathbb E_0T_{Nx}\geq xN^2/(2\alpha\chi)$ for a target fraction $x>0$. The quadratic scaling reflects the decreasing separation between adjacent transition frequencies: maintaining their resolution constrains the absolute jump rates. For fixed-pair coupling, the corresponding bound grows linearly with the target occupation over the finite-size range in which the ladder remains ascending.
	
\begin{figure}[t]
		\centering
		\includegraphics[width=\columnwidth]{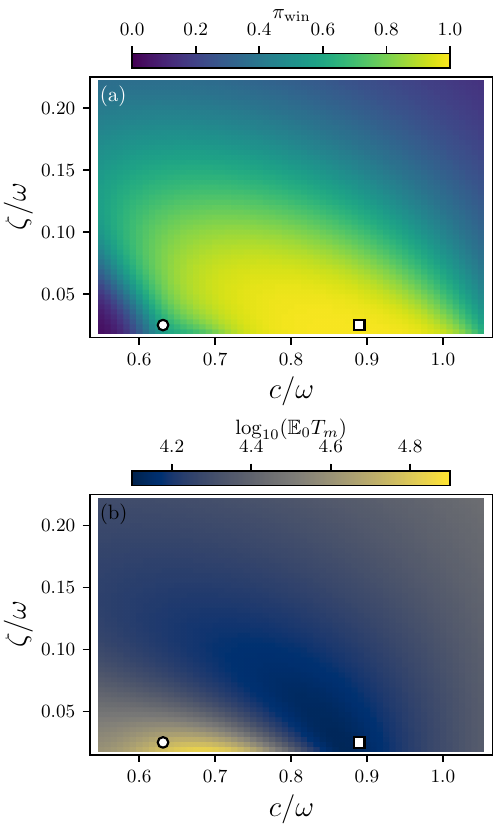}
		\caption{Target localization and first-passage dynamics for the cutoff family of Fig.~\ref{fig:stationary_map}. Panel (a) gives $\pi_{\rm win}=\sum_{|k-8|\leq3}\pi_k$, the stationary probability within three excitations of the target level $m=8$. Panel (b) shows $\log_{10}(\mathbb E_0T_8)$, the base-10 logarithm of the mean first-passage time from $k=0$ to $k=8$. The axes show the normalized cutoff center $c/\omega$ and spectral width $\zeta/\omega$. Circle and square mark the same equal-mean, equal-budget reservoirs as Fig.~\ref{fig:stationary_map}. A common weak-rate reduction enforces the transition-resolution ceiling and sets the absolute time scale.}
		\label{fig:kinetics_map}
	\end{figure}

	Figure~\ref{fig:kinetics_map} compares stationary occupation near the target with the time needed to reach it from the empty battery. Shifting the cutoff at fixed mean charge, width, and upward-rate capacity both concentrates the stationary population near the target and increases traffic through the preceding edges. The resulting shorter first-passage time and stronger localization arise from different portions of the same spectral profile (Table~\ref{tab:center}).
	
	\begin{table}[t]
		\caption{Stationary charge at equal mean occupation and upward-rate capacity ($N=20$, $\langle k\rangle/N\simeq0.42$). The target window is $|k-8|\leq3$, and $T_8=\mathbb E_0T_8$ is the first-passage time from the empty battery. Changes are measured relative to the lower-center case (pp: percentage points); the matched operating parameters are specified in Appendix~\ref{app:numerical-reservoirs}.}\label{tab:center}
		\begin{ruledtabular}
			\begin{tabular}{lrrr}
				& Lower & Upper & Change\\
				\colrule
				$c$ & $0.631$ & $0.890$ & ---\\
				$R_{\rm CSS}$ & $4.08$ & $0.292$ & $-93\%$\\
				$\pi_{\rm win}$ & $58.9\%$ & $98.9\%$ & $+40.0$ pp\\
				$\erg_\sym/N$ & $0.130$ & $0.265$ & $+104\%$\\
				$T_8/T_8^{\rm L}$ & $1$ & $0.264$ & $-74\%$\\
			\end{tabular}
		\end{ruledtabular}
	\end{table}
	
	For the upper-center pump, transitions below $m$ carry population toward the target, while a falling pump-to-loss ratio above it restricts further excitation. The lower-center pump leaves a broader high-occupation tail and slower arrival at $m$. This separation of access and retention extends the role of frequency-selective collective transitions in charging and superabsorption~\cite{Higgins2014,Pokhrel2025,AlvarezCuartasReina2026}, and also occurs in the tested fixed-pair window. During repeated operation, the initial distribution for each charging contact is set by the preceding extraction stroke.
	
	\subsection{Preparation-dependent recharge}\label{sec:mpemba}
	
	The preparation entering a charging contact after extraction generally differs from the empty state used for the first-passage comparison. The same birth--death generator then relaxes a different population distribution toward $\bm\pi$. For positive neighboring rates, the generator is reversible and has real relaxation eigenvalues. If $\phi_1(k)$ is a left eigenfunction of its slowest nonstationary mode, the corresponding amplitude is
	\begin{equation}\label{eq:slow}
		a_1(\bm p)=\sum_k\phi_1(k)[p_k-\pi_k],
		\qquad L^T\bm\phi_1=-\lambda_1\bm\phi_1.
	\end{equation}
	The eigenfunction can be oriented to increase strictly with occupation. A stochastic redistribution that only moves population downward from $\bm\pi$ then has $a_1<0$ whenever the redistribution is nontrivial. The selected $\ket0\leftrightarrow\ket m$ work operation acts in both directions, and its actual post-stroke distribution is used in the recharge calculation.
	
	At the upper-center operating point, a diagonal preparation can be chosen with $a_1=0$, while preserving the reference post-stroke mean energy, sector ergotropy, and population ordering. This is obtained within the normalized, positive distributions by imposing $\bm\phi_1^T(\bm p-\bm\pi)=0$ along with the same energy and ergotropy constraints. The charging generator, contact duration, and subsequent extraction stroke are then kept fixed.
	
	The resulting strong-Mpemba preparation approaches the stationary distribution more closely in total variation at an equal scaled recharge time, yet yields slightly less mean work in the next extraction. Canceling the slowest relaxation mode accelerates convergence toward $\bm\pi$, whereas the specified work stroke responds to the occupations it exchanges. The distinction between relaxation and recoverable work is also pertinent to dissipative quenches of quantum batteries~\cite{Medina2025,Grazi2026DissipativeQuenches}. 
	
	\section{Extracted work and long-time reliability}\label{sec:work}
	
	The charge recovered between successive extraction strokes determines both the next work event and its dependence on previous events. In a finite contact, the charging propagator transfers population from the post-stroke state toward the charged stationary distribution. The resulting sequence of work outcomes is correlated, so its accumulated variance contains information about the transport during every recharge. The counting statistics introduced in Sec.~\ref{sec:model} provide a direct way to quantify this connection~\cite{Mohan2026,Rinaldi2026}. Related collective Otto cycles exhibit contact-induced work correlations under passive reservoir shaping~\cite{DamasNeto2026ThermodynamicRouting}. For the battery, each cycle instead counts the energy delivered by a selective extraction stroke following active recharge.
	
	To isolate the influence of spectral placement on repeated-work fluctuations, we compare two charging reservoirs coupled to the same $\ket0\leftrightarrow\ket8$ cyclic operation. Their charging contacts are adjusted to deliver the same mean work in their respective periodic states. This cyclic matching probes the population redistribution between successive extractions; the equal-mean-occupation comparison of Sec.~\ref{sec:stationary} instead characterizes stationary charging. Table~\ref{tab:work} summarizes how the two protocols differ in fluctuations, energy throughput, and charging duration at this matched output.
	
	\begin{figure}[t]
		\centering
		\includegraphics[width=\columnwidth]{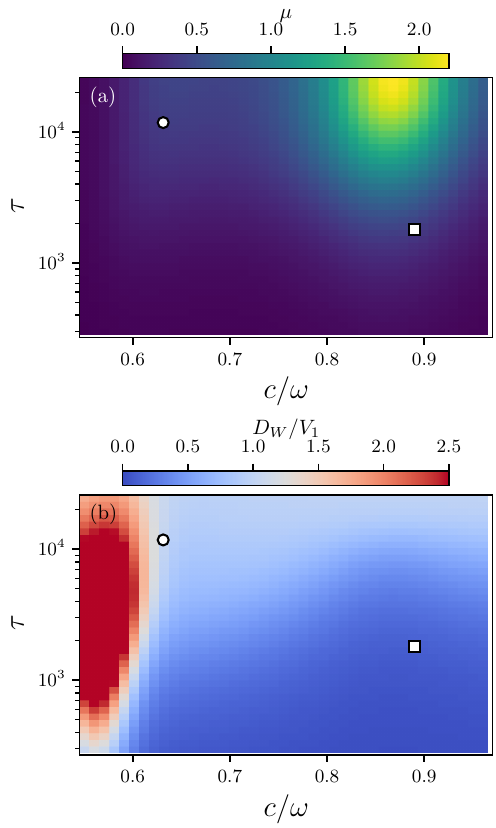}
		\caption{Mean extracted work and temporal reliability of the cyclic battery for the ideal $\ket0\leftrightarrow\ket8$ exchange, $N=20$, and cutoff width $\zeta=0.025$. Panel (a) shows the mean extracted two-point-measurement work per cycle $\mu$, and panel (b) the ratio $D_W/V_1$ of asymptotic variance growth per cycle to single-cycle work variance. The axes are cutoff center $c/\omega$ and charging-contact duration $\tau$. The circle and square mark protocols retimed to deliver the same mean work in their periodic states. Table~\ref{tab:work} gives the corresponding variance and covariance comparisons. The same reservoir rate ceiling applies throughout.}
		\label{fig:work_map}
	\end{figure}
	
	\begin{table}[t]
		\caption{Work statistics for the ideal selective exchange at matched mean work ($\mu\simeq0.395$). Superscript L denotes the lower-center reference; $D_W/V_1-1$ gives the normalized sum of intercycle covariances, $\Pi_\tau=\mu^2/(D_W\tau)$ is precision per charging-contact time, and $\mu/E_+$ measures recovery relative to gross pumped energy. The change column gives relative differences or percentage points (pp).}\label{tab:work}
		\begin{ruledtabular}
			\begin{tabular}{lrrr}
				& Lower & Upper & Change\\
				\colrule
				$\tau/\tau^{\rm L}$ & $1$ & $0.153$ & $-85\%$\\
				$V_1/V_1^{\rm L}$ & $1$ & $4.86$ & $+386\%$\\
				$D_W/D_W^{\rm L}$ & $1$ & $0.541$ & $-46\%$\\
				$D_W/V_1-1$ & $+5.0\%$ & $-88.3\%$ & ---\\
				$\Pr(w>0)$ & $6.94\%$ & $19.4\%$ & $+12.5$ pp\\
				$\Pr(w<0)$ & $0.0056\%$ & $12.5\%$ & $+12.5$ pp\\
				$\mu/E_+$ & $12.9\%$ & $52.0\%$ & $+39.1$ pp\\
				$\Pi_\tau/\Pi_\tau^{\rm L}$ & $1$ & $12.1$ & $12.1\times$\\
			\end{tabular}
		\end{ruledtabular}
	\end{table}
	
	Figure~\ref{fig:work_map} and Table~\ref{tab:work} reveal a reversal in work fluctuations at matched mean output. The upper-center reservoir produces nearly five times the variance in an individual stroke, yet the variance of accumulated work grows at roughly half the rate. A negative relaxation mode of the charge--extract map generates alternating correlations whose integrated contribution suppresses long-time fluctuations. The precision of repeated work delivery therefore depends on how the charging spectrum redistributes population between extractions, not only on the probabilities of individual work outcomes.
	
	The larger single-cycle variance of the upper-center protocol reflects appreciable probabilities of both work extraction and work investment; investment events are rare for the lower-center pump. After each exchange, the resulting occupations seed the next recharge. The sign and strength of the work correlations are therefore set by the full charge--extract map rather than by the width of a single-stroke distribution.
	
	Figure~\ref{fig:work_map} separates average output from temporal reliability across the cutoff-control landscape. Panel (a) identifies the mean-work contours along which the periodic protocols can be matched; panel (b) reveals the effect of work correlations through $D_W/V_1$. Ratios below unity signify that temporal anticorrelations suppress the variance accumulated across many strokes. The variance reversal persists for tested alternative two-level exchanges, a second cutoff width, and a finite fixed-pair interaction (Appendix Fig.~\ref{fig:app_swap}). Within a sampled family of single-Lorentzian pumps satisfying the same rate budget and resolution constraint, the selected cutoff achieves a smaller long-time variance and roughly a threefold advantage in charging-contact-time-normalized precision. The variance ordering is consequently a property of the reservoir spectrum together with the extraction operation and contact duration. A separate calibration scan of the upper-center cutoff confirms that the long-time variance advantage persists under bounded shifts of the spectral center and width, after retiming to equal mean work (Appendix~\ref{app:valid}).
	
	The shorter charging contact gives the upper-center protocol about twelve times the accumulated-work precision per charging-contact time. Including the work pulse and any isolated storage interval changes the elapsed cycle time (Sec.~\ref{sec:model}). At periodic stationarity the mean extracted work satisfies $\mu=E_+-E_-$, where $E_+$ and $E_-$ are the gross energies entering and leaving the battery through the two reservoirs. Approximately half of the pumped energy is recovered as work by the upper-center protocol, compared with about one-eighth for the lower-center protocol (Table~\ref{tab:work}). These are pump-channel recovery fractions; a device-level efficiency also includes the resources required to sustain the pump and operate the controller.

	\subsection{Temporal memory and energy counting}
	
	The exchange operation provides a simple limit in which to isolate the origin of correlations. With vanishing charging duration, two successive exchanges restore the original occupations, and their work contributions cancel pathwise. The accumulated work over each pair of strokes is then zero. At finite charging duration, population redistribution between exchanges modifies this cancellation. The spectral response controls the relative occupation transfers during recharge and thereby determines the remaining correlation between successive work outputs.
	
	Work counting is also related exactly to energy transport into the battery. Let $b_{j-1}$ be the battery energy immediately after the preceding extraction, $a_j$ its energy after charging, and $b_j$ the energy after the next extraction. The net battery-energy change during contact $j$ is $q_j=a_j-b_{j-1}$, and the extracted work is $w_j=a_j-b_j$. Summing over $M$ cycles gives
	\begin{equation}\label{eq:telescoping}
		\sum_{j=1}^M w_j=\sum_{j=1}^M q_j-(b_M-b_0).
	\end{equation}
	The initial-to-final stored-energy difference stays bounded because the battery has finitely many levels. Over long sequences, the work delivered by extraction consequently records the net energy replenished during the charging contacts, with identical scaled cumulant-generating functions and cumulants per cycle. Appendix~\ref{app:count} establishes the equivalent similarity relation between their tilted-cycle maps. Gross pumping energy $E_+$ includes the energy later removed through the loss channel; the ratio $\mu/E_+$ in Table~\ref{tab:work} summarizes its contribution to the recovered work.
	
	The sign and overlap of the charge--extract relaxation modes explain the different covariance sums. The upper-center map has a negative mode that alternates the sign of successive correlations and dominates the integrated suppression. The lower-center map contains a slower positive mode with contributions partially offset by faster modes. Their quantitative decomposition is given in Appendix~\ref{app:count}.
	
	The correlations between repeated work outputs thus belong to the dynamics of energy replenishment, complementing studies of quantum-battery fluctuations and the statistics of repeated operation~\cite{Mohan2026,Rinaldi2026,GarciaPintos2020Fluctuations,Bakhshinezhad2024,Barra2022Cycle}.
	
	\section{Symmetry leakage and finite-use work}\label{sec:leakage}
	
	Local dephasing preserves excitation number but redistributes population among sectors of different total spin. The fully symmetric Dicke ladder contains one state for each $k$, whereas the full ensemble has $\binom Nk$ states at the same excitation number. A collective work operation preserves total spin, so this redistribution changes the work accessible to the controller even when stored excitation remains. Related distinctions between stored energy and accessible ergotropy arise in locally restricted and noisy charging protocols~\cite{Ali2026AQTAccessibleWork,Grazi2026AQTFiniteTime}.
	
	We probe this effect with a collective transverse $J_x$ pulse applied within every spin sector. It differs from the ideal selective exchange used in Sec.~\ref{sec:work} to isolate the correlations between repeated work events. Charging is described by a permutation-invariant generator with frequency-resolved collective pump and loss and independent local dephasing $\gamma_\phi\sum_i\Diss[\sigma_z^{(i)}]$~\cite{Shammah2018,Canzio2025}. For $N=20$, the reduced population dynamics contain $121$ distinct $(J,m)$ coordinates.
	
	The common spectral functions across spin sectors follow from the collective form of the coupling. Since $H_B$ depends only on $J_z$, the transition $\ket{J,m}\to\ket{J,m+1}$ has frequency $\Omega_k=\omega+\lambda_N(2k-N-1)$ with $k=m+N/2$, independent of $J$. The homogeneous operator $J_+$ therefore samples $S_+(\Omega_k)$ on every such transition, while its squared matrix element $C_{J,m}^{+}=(J-m)(J+m+1)$ sets the sector-dependent rate; the analogous statement holds for loss through $J_-$. The same spectral ratios therefore enter every spin sector, but with sector-dependent transition strengths. Local dephasing connects the sectors, producing a stationary state that can be found exactly. Define $a_0=1$, $a_k=\prod_{j=1}^k r(\Omega_j)$, and let $\Pi_k$ project onto the complete $k$-excitation subspace. For positive collective edge rates and any $\gamma_\phi>0$, continuous charging with local dephasing gives
	\begin{align}\label{eq:leakage_stationary}
		Z_N&=\sum_{k=0}^N\binom Nk a_k,\nonumber\\
		\rho_\infty&=\frac{1}{Z_N}\sum_{k=0}^N a_k\Pi_k,\nonumber\\
		p_{\rm sym}^{(\infty)}&=\frac{1}{Z_N}\sum_{k=0}^N a_k.
	\end{align}
	The collective jumps balance between neighboring excitation manifolds, while the dephasing channel mixes their total-spin content. The excitation-number probabilities are $P_k=\binom Nk a_k/Z_N$ and hence satisfy
	\begin{equation}\label{eq:leakage_manifold_ratio}
		\frac{P_k}{P_{k-1}}=\frac{N-k+1}{k}\,r(\Omega_k).
	\end{equation}
	The factor $\binom Nk$ counts the states sharing excitation number $k$. It multiplies the spectral weight $a_k$ in the full ensemble, distributing occupation over many spin sectors while each $k$ contains only one fully symmetric Dicke state. Spectral selection can therefore keep the total excitation concentrated even as the population accessible through the maximal-spin sector becomes small. For both cutoff profiles the resulting stationary population of the maximal-spin sector is of order $10^{-5}$ in the representative twenty-spin system~\cite{Khoudiri2026,ZhangJing2025StableCharging}. Its limiting value is determined by the reservoir spectrum and excitation-manifold multiplicities; the dephasing strength sets the timescale on which this distribution is approached. In the permutation-invariant population dynamics, the lowest nonzero relaxation rate scales approximately linearly with weak local dephasing. The corresponding inverse-gap times are much longer than a single charging contact at the representative twenty-spin operating point. This separates the finite-use loss of collective accessibility from the asymptotic symmetry-diluted stationary state (Appendix~\ref{app:local_robustness}). Periodic extraction produces a different stationary cyclic state.
	
	To evaluate work after population has spread between sectors, we apply a collective transverse pulse generated by $H_B+\Omega_xJ_x$ to the whole spin ensemble. The pulse acts within each $J$ sector, and its energy-basis transition probabilities define the work stroke for two-point measurements. The controller parameters and charging-contact times are fixed by a common clean reference, with equal mean extracted work for both reservoirs. Under repeated operation, the total extracted work $W_K=\sum_{n=1}^K w_n$ has finite-horizon precision
	\begin{equation}\label{eq:finite_horizon_precision}
		\Pi_K=\frac{\langle W_K\rangle^2}{\Var(W_K)}.
	\end{equation}
	We evaluate the finite-$K$ counting statistics with the energy-basis transition probabilities of the same transverse pulse at each cycle. This connects the accumulated work precision to the actual redistribution of charge across spin sectors.  For compact figure notation we also write the accumulated mean extracted work per cycle as $\mu_K\equiv \langle W_K\rangle/K$. For this collective pulse, the upper-center protocol already has slightly smaller single-cycle variance in the clean model, and its intercycle correlations further reduce the long-time variance. The reversal from larger single-cycle variance to smaller long-time variance remains specific to the selective $\ket0\leftrightarrow\ket m$ exchange.
	
	\begin{figure}[t]
		\centering
		\includegraphics[width=\columnwidth]{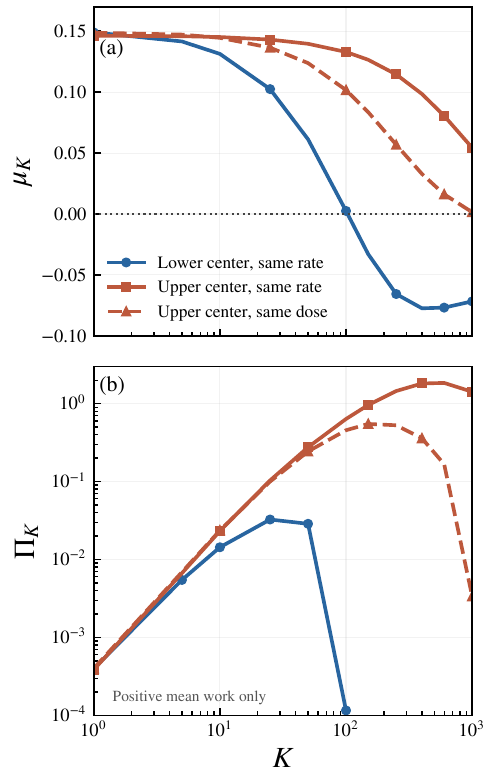}
		\caption{Finite-use work delivery and precision under independent local-spin dephasing at $\gamma_\phi/\Gamma_{\rm res}=10^{-4}$, with $\Gamma_{\rm res}=0.04\Delta\Omega$. (a) Accumulated mean extracted TPM work per cycle, $\mu_K\equiv \langle W_K\rangle/K$, for the lower-center and upper-center chargers, using the same physical per-spin dephasing rate (solid lines). The dashed line increases the upper-center rate to match the lower charger's integrated dephasing dose per charging contact. (b) Finite-horizon precision $\Pi_K=\langle W_K\rangle^2/\Var(W_K)$, plotted where $\mu_K>0$. All three cases use the same sector-wide transverse work pulse, $H_B+J_x$, for a duration $t_{\rm w}=1.5$; each begins in its clean periodic post-extraction state. The charging-contact durations were matched for equal clean mean work, and local noise acts during charging. Symbols mark computed cycle counts; connecting lines guide the eye.}
		\label{fig:local_noise_window}
	\end{figure}
	
	The finite-use advantage becomes visible when accumulated work and its precision are considered together. Figure~\ref{fig:local_noise_window}(a) follows the mean work delivered during repeated charge--extract cycles in a common dephasing environment. The lower-center charger crosses into net work investment after roughly one hundred cycles, whereas the upper-center protocol continues to produce positive mean output over the sampled horizon. Panel (b) shows the precision of accumulated output while the corresponding mean work remains positive. The same-rate comparison includes the shorter charging contact of the upper-center reservoir, which reduces its exposure to local noise per cycle.
	
	The dashed upper-center curve provides a complementary comparison: its dephasing rate is increased until both chargers experience the same integrated dephasing dose per contact. After one hundred cycles, their maximal-spin populations are nearly equal, approximately $6$--$7\%$, yet the upper-center protocol still delivers appreciable mean work and greater accumulated-work precision. The difference reflects the distribution of charge among the accessible sectors and the subsequent recharge dynamics, rather than a larger surviving symmetric fraction. Precisions associated with $\mu_K<0$ are excluded from Fig.~\ref{fig:local_noise_window}(b), since they quantify reproducible net investment rather than reliable extraction.
	
	Equal physical elapsed time provides a second operational comparison. The shorter-contact reservoir completes more cycles and delivers greater accumulated work and precision within the tested window. Contact-time scans also contain operating points where the lower-center charger has better precision per elapsed time. The advantage is consequently tied to the chosen spectral placement, operating contact, and transverse work controller, rather than a universal ordering of the reservoirs.
	
	Over sufficiently many cycles, weak local processes redistribute the population among the available spin sectors, and the tested collective controller eventually requires mean invested work for both reservoirs. Finite-use positive output and the dephasing-controlled stationary limit consequently characterize different operating regimes. Appendix~\ref{app:local_robustness} gives the sector transitions, stationary-state derivation, and counting procedure, together with the full-spin disorder-during-charging benchmark.

	\section{Discussion and conclusions}
	\label{sec:conclusion}
	
	In an interacting Dicke battery, the competition between frequency-selective pumping and loss governs both the charge that accumulates and the routes by which it can be replenished. The collective matrix elements cancel from the stationary population ratios but remain in the transition rates. Spectral placement can therefore sharpen the charged distribution, increase its ergotropy, and shorten the time to reach a selected excitation at fixed mean occupation and upward-rate capacity.
	
	The system-size dependence exposes two distinct charging regimes. At low excitation, the extensive-coupling limit becomes harmonic and its stationary state is passive. At finite excitation density, the interaction continues to separate the frequencies sampled by the occupied ladder, and the local slope of the pump-to-loss ratio controls fluctuations around the favored charge. Resolving those transitions as their spacing decreases also determines how the access time scales with system size. The corresponding charge--extract correlations belong to that same scaling of rates and contact durations.
	
	During repeated extraction, the same rates become the source of temporal work correlations. For the selective $\ket0\leftrightarrow\ket m$ operation, the reservoir with the larger variance in a single stroke has the smaller long-time variance at matched mean work because its finite-recharge dynamics generate negative intercycle covariance. The counting identity relates the accumulated work to net charging-energy transfer up to the bounded change in stored battery energy. The slow modes governing approach to the stationary state also need not govern recovery of work through the selected operation.
	
	With local dephasing, excitation spreads among total-spin sectors, and their multiplicities lead to an exact stationary symmetry-dilution law. A collective transverse controller samples this redistributed charge differently from the ideal exchange and sustains positive mean work with finite-horizon precision over a limited operating interval. Spectral selection thus determines not only how charge is prepared but how reliably a specified controller can recover it during repeated use.
	
	In a microscopic implementation, the spectral response of the powered reservoir, the energy required to sustain it, and the dynamics of work extraction are connected through the device architecture. Resolving these contributions would allow reservoir profiles to be compared at fixed total energetic expenditure, complementing the matched transition-rate and mean-work comparisons developed here. Increasing the system size introduces another physical scale: for extensive coupling, neighboring Dicke transitions approach one another with spacing $2\chi/N$. When this separation becomes comparable to the dissipative linewidths, partially resolved dynamics can couple neighboring transitions and modify both charge preparation and temporal work correlations. A joint treatment of the reservoir spectrum and a finite-duration collective work controller would then establish how spectral selectivity can be used to improve accumulated-work precision under a common energy and time budget. Such an approach would connect the present analytical results to the design of complete, repeatedly operated quantum-battery devices.

	\begin{acknowledgments}
		H.-G.X. acknowledges the support from Start-up Fund of the School of Physics and Optoelectronic Engineering, Guangdong Polytechnic Normal University, under Grant No. 2026SDKYA018.
		OpenAI ChatGPT assisted with scientific discussion and analytical cross-checks, and drafting and revision of the manuscript. The authors directed these uses, independently checked the scientific outputs, reviewed and edited the final content, and take full responsibility for the results and manuscript. 
	\end{acknowledgments}
	
	\section*{Data and code availability}
	Data and numerical code supporting the findings of this study are available from the corresponding author upon reasonable request.
	\appendix
	\section{Interaction continuations and numerical reservoir parameters}\label{app:numerical-reservoirs}
	
	The finite-size comparisons use $\omega=1$, $\chi=0.48$, and $\lambda_0=0.08$. Thus $\lambda_N=\chi/N$ and $\lambda_N=\lambda_0$ coincide at $N_0=6$ but define distinct continuations thereafter. The fixed-pair model has positive ascending gaps only through $N=13$ with these fixed parameters. The reported results outside this window use the extensive prescription, not an extrapolation of fixed pair coupling.
	
	For each chosen profile $f$ from Eq.~\eqref{eq:cutoff}, the unscaled upward reservoir and the fixed loss reservoir in Figs.~\ref{fig:stationary_map}--\ref{fig:work_map} are
	\begin{align}\label{eq:reference_spectra}
		S_{-,0}(\Omega)&=0.034+0.009\frac{0.27^2}{(\Omega-0.93)^2+0.27^2},\\
		S_{+,0}^{(f)}(\Omega)&=\frac{B_{N,0}\,f(\Omega)}{\sum_{j=1}^N C_j f(\Omega_j)},\nonumber\\
		B_{N,0}&=0.065\sum_{j=1}^N C_j.\nonumber
	\end{align}
	The floor in the cutoff and single-Lorentzian profiles is $b=0.06$. This normalization fixes the \emph{unscaled} total upward edge-rate capacity and not the pump's thermodynamic free-energy consumption. A frequency-flat reference separately retuned to match a chosen stationary mean occupation generally has a different rate budget.
	
	In every computed figure the propagated rates are $u_{k-1}=s_N C_k S_{+,0}^{(f)}(\Omega_k)$ and $d_k=s_N C_k S_{-,0}(\Omega_k)$. The effective budget is therefore
	\begin{equation}\label{eq:B}
		B_N\equiv\sum_{k=1}^N u_{k-1}=s_N B_{N,0}.
	\end{equation}
	The multiplier $s_N$ is common to the reservoir profiles compared at fixed system size and rescales all contact durations inversely. For the principal extensive-interaction data at $N=20$, the choice $s_{20}\simeq8.07\times10^{-5}$ ensures the resolved-edge ceiling $\max_k(u_k+d_k)/\Delta\Omega\leq0.04$, with $\Delta\Omega=2|\lambda_N|$. At the exactly degenerate point $\lambda_N=0$, transitions share a Bohr frequency and collective jump grouping becomes relevant to the full quantum dynamics.

	\paragraph*{Matched reference settings.}
	The two cutoff centers are $c_{\rm L}=0.631364$ and $c_{\rm U}=0.889755$. For the selective exchange, the matched charging durations are $\tau_{\rm L}=11765$ and $\tau_{\rm U}=1803$, giving a common mean extracted work $\mu\simeq0.395$. These are the operating settings underlying Tables~\ref{tab:center} and \ref{tab:work}.
	
	\section{Independent-channel population generator and stationarity}\label{app:generator}
	
	An energy-resolved Lindblad description leading to Eq.~\eqref{eq:rates} is
	\begin{align}
		\dot\rho={}&-i[H_B,\rho]\nonumber\\
		&+\sum_{k=1}^N S_+(\Omega_k)
		\Diss[\sqrt{C_k}\ket{k}\bra{k-1}]\rho\nonumber\\
		&+\sum_{k=1}^N S_-(\Omega_k)
		\Diss[\sqrt{C_k}\ket{k-1}\bra{k}]\rho.
	\end{align}
	where $\Diss[L]\rho=L\rho L^\dagger-\{L^\dagger L,\rho\}/2$. The rate equation is diagonal-invariant under this generator. At stationarity, each edge carries zero \emph{net population current} $u_{k-1}\pi_{k-1}=d_k\pi_k$, giving Eq.~\eqref{eq:pi}. This does not imply zero \emph{gross energy throughput} when the upward and downward processes are maintained by distinct physical reservoirs. Multiplying every rate by the same positive number rescales time without changing $\bm\pi$, while changing the spectral ratios can reshape $\bm\pi$. For $\chi=0$, all transition frequencies coincide, and spectral-shape changes reduce to amplitude retiming at the population level; coherence dynamics require separate attention because the degenerate quantum jump operators can group collectively.
	
	The macro-limit Eq.~\eqref{eq:ldp} follows from $N^{-1}\sum_{j=1}^{Nx}\ln r(\Omega_j)$ as a Riemann sum. At a smooth interior maximum $\Phi'(x_*)=\ln r(\Omega(x_*))=0$, with $\Phi''(x_*)=2\chi\partial_\Omega\ln r(\Omega_*)<0$, the local variance is $\Var(k)\sim -N/\Phi''(x_*)$. For distributions concentrated on an $O(\sqrt N)$ window about $Nx_*$, passive sorting packs the large eigenvalues near the bottom of the ascending ladder; its energy is $o(N)$, which yields the stated sector-ergotropy density. These statements do not cover multiple coexisting stationary maxima or singular filter profiles without modification.
	
	\begin{figure}[t]
		\centering
		\includegraphics[width=0.49\textwidth]{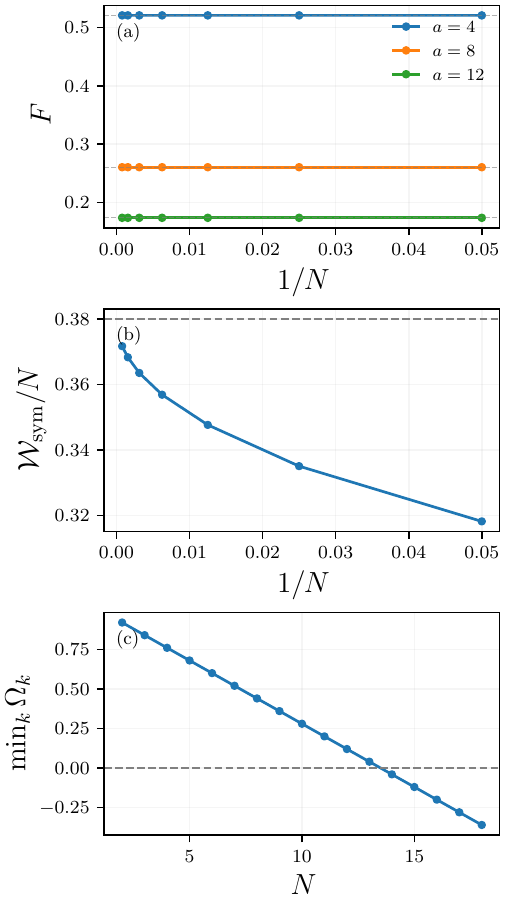}
		\caption{Finite-size and interaction-continuation controls. (a) Fano factor $F=\Var(k)/\langle k\rangle$ for the exactly solvable extensive-coupling family with reservoir ratio $r(\Omega)=\exp[-a(\Omega-\omega)]$ at several slopes $a$, versus $1/N$; dashed lines are $F_\infty=(a\chi)^{-1}$. (b) Symmetric-sector ergotropy density $\mathcal W_{\rm sym}/N$ for $a=8$ and its asymptote $\omega/2-\chi/4$ (dashed line), with $\omega=1$ and $\chi=0.48$. (c) The smallest adjacent gap $\min_k\Omega_k$ in the fixed-pair $\lambda_0=0.08$ continuation at fixed $\omega=1$; the zero line identifies the loss of a strictly ascending ladder. The analytic exponential-ratio family is a limiting-case check, not a fit to the cutoff map.}
		\label{fig:app_largeN}
	\end{figure}
	
	\section{First-passage recursion and the resolution-dependent bound}\label{app:mfpt}
	
	Let $t_i$ be the expected hitting time of $m$ from $i<m$. The backward equation is $-1=u_i(t_{i+1}-t_i)+d_i(t_{i-1}-t_i)$, with $t_m=0$ and $d_0=0$. Setting $\delta_i=t_i-t_{i+1}$ gives $u_i\delta_i=1+d_i\delta_{i-1}$, with $\delta_0=1/u_0$, proving Eq.~\eqref{eq:mfpt}. The result is independently reproduced by inversion of the absorbing subgenerator.
	
	For the lower bound, let $J_{T_m}$ be the number of jumps before the first hitting time. Nearest-neighbor motion from $0$ to $m$ implies $J_{T_m}\ge m$ pathwise. If each state's escape intensity $R_k$ satisfies $R_k\le\Gamma_{\max}$, the compensator identity gives $m\le\mathbb E J_{T_m}=\mathbb E\int_0^{T_m}R_{X_t}\,dt\le\Gamma_{\max}\mathbb ET_m$. The additional condition $\Gamma_{\max}\le\alpha\Delta\Omega$ is an explicit resolved-transition \emph{model assumption}. It becomes restrictive as $\Delta\Omega\propto N^{-1}$ in the extensive continuation. No such $N\to\infty$ extrapolation is made at fixed pair interaction and fixed $\omega$, because the ascending energy spectrum has already ceased to exist.
	
	\section{Counting-field identities, covariance, and numerical checks}\label{app:count}
	
	Let $T_1=D\diag(w)C_\tau$, $T_2=D\diag(w^2)C_\tau$, and $P=D C_\tau$. For the stationary post-stroke distribution $\bm q$, write $\mu=\bm1^TT_1\bm q$, $V_1=\bm1^TT_2\bm q-\mu^2$, and $\bm v=T_1\bm q-\mu\bm q$. Solving
	\begin{equation}
		(I-P+\bm q\bm1^T)\bm y=\bm v
	\end{equation}
	provides the exact covariance correction
	\begin{equation}
		D_W=V_1+2\bm1^TT_1\bm y.
	\end{equation}
	For the mixing cycle maps considered here, the covariance sum, finite-cycle moment recursion, and tilted-eigenvalue derivatives give the same asymptotic variance. The linear system avoids explicit summation over arbitrarily many cycles.
	
	The correlation mechanism can also be decomposed into nonstationary modes of $P$. Define $\bm h^T=\bm1^TT_1$ and $\bm v=T_1\bm q-\mu\bm q$. For a diagonalizable cycle map, let $P\bm r_a=\lambda_a\bm r_a$ and $\bm l_a^TP=\lambda_a\bm l_a^T$, with $\bm l_a^T\bm r_b=\delta_{ab}$. Then
	\begin{align}\label{eq:work_mode_covariance}
		\operatorname{Cov}(w_0,w_s)&=\sum_{a\ne0}A_a\lambda_a^{s-1},\qquad s\ge1,\nonumber\\
		D_W-V_1&=2\sum_{a\ne0}\frac{A_a}{1-\lambda_a},\nonumber\\
		A_a&=(\bm h^T\bm r_a)(\bm l_a^T\bm v).
	\end{align}
	The stationary mode does not contribute because $\bm1^T\bm v=0$; complex-conjugate modes contribute real sums. For the upper-center protocol a mode near $-0.45$ supplies most of the negative integrated covariance (about $85\%$). The lower-center protocol has a slower positive mode near $0.91$, whose contribution is partly canceled by faster modes. The same sign contrast appears at a second tested cutoff width. These amplitudes depend jointly on the charging propagator and the work operation, explaining why spectral placement can reverse the long-time variance ordering.
	
	For completeness, let $C_{ij}$ be the conditional charging probability from post-stroke level $j$ to pre-stroke level $i$, and let $D_{ki}$ be the probability of the extracted post-stroke level $k$ given $i$ (a permutation in the main examples). The matrix generating \emph{extracted work} is
	\begin{equation}
		(T_W)_{kj}(s)=\sum_i D_{ki}C_{ij}\,e^{s(\epsilon_i-\epsilon_k)},
	\end{equation}
	while the generator for the \emph{net charging-energy change} is
	\begin{equation}
		(T_Q)_{kj}(s)=\sum_i D_{ki}C_{ij}\,e^{s(\epsilon_i-\epsilon_j)}.
	\end{equation}
	With $F_s=\diag(e^{s\epsilon_0},\ldots,e^{s\epsilon_N})$,
	\begin{equation}\label{eq:similarity}
		\boxed{T_W(s)=F_{-s}T_Q(s)F_s.}
	\end{equation}
	Their spectra and asymptotic cumulants therefore coincide. This is an exact similarity relation for the specified energy-resolved cyclic maps and remains valid for general stochastic discharge kernels $D$ with outcome-resolved energy changes. The associated pathwise identity is Eq.~\eqref{eq:telescoping}. Neither relation implies equality of extracted work with the gross energy supplied by the powered pump.
	
	\begin{figure}[t]
		\centering
		\includegraphics[width=0.49\textwidth]{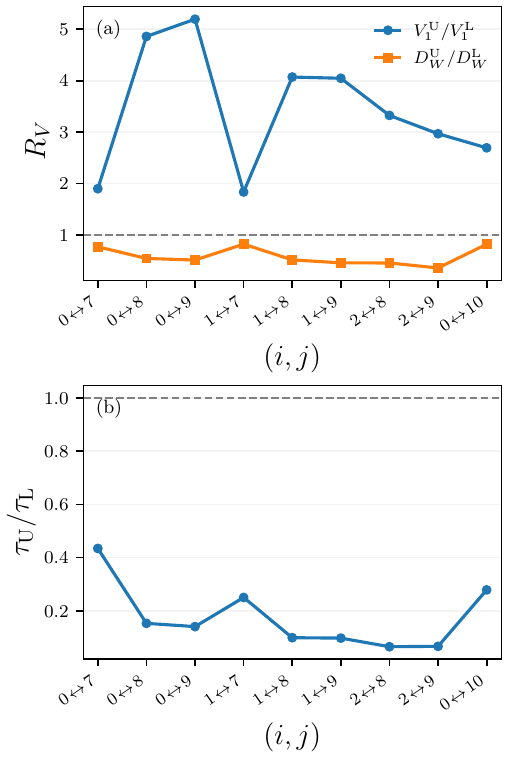}
		\caption{Dependence of the repeated-work comparison on the chosen cyclic two-level extraction stroke for the $N=20$ extensive-interaction battery. Each point uses a common swap $\ket i\leftrightarrow\ket j$ and retimes the upper-center reservoir to match the lower-center mean extracted work. (a) The vertical metric $R_V$ denotes the upper-to-lower variance ratio, plotted separately for one-cycle variance $V_1$ and long-time variance growth $D_W$; the dashed line is equality. (b) Matched-work contact-time ratio $\tau_{\rm U}/\tau_{\rm L}$ (dashed line: equal duration). The controls demonstrate persistence within nine tested ideal swaps, not optimization over arbitrary work operations.}
		\label{fig:app_swap}
	\end{figure}
	
	For the single-Lorentzian comparison, the sampled grid contains 19 center frequencies $c/\omega\in[0.45,1.05]$ and widths $\zeta/\omega\in\{0.015,0.025,0.04,0.06,0.10,0.16,0.25\}$. The upward-rate budget and resolution criterion are those of the cutoff comparison. Feasible cases are retimed to the same mean work, $\mu\simeq0.395$.
	
	\begin{figure}[t]
		\centering
		\includegraphics[width=0.49\textwidth]{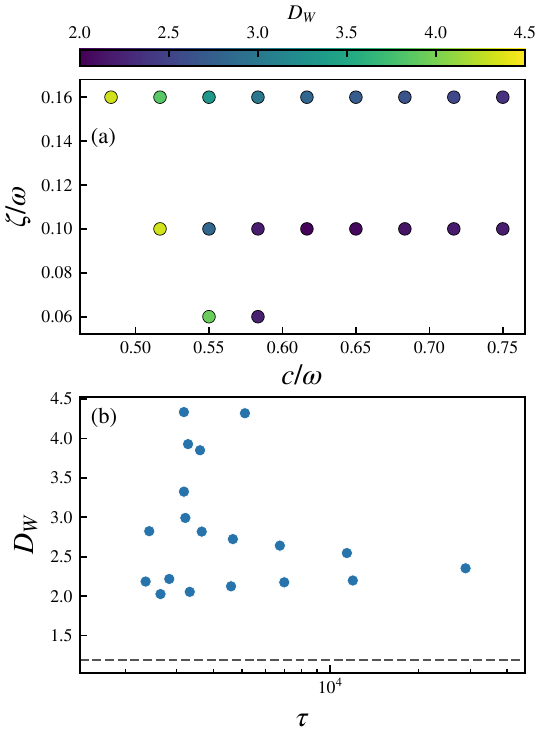}
		\caption{Repeated-work comparison with a bounded single-Lorentzian pump family at $N=20$. (a) Long-time work-variance growth $D_W$ for centers and widths that satisfy the rate-resolution ceiling and admit equal-mean-work retiming. (b) $D_W$ against the corresponding charging-contact duration for the common weak-rate continuation $\eta=0.85$; the dashed line marks the upper-center cutoff reference, $D_W\simeq1.19$. Of the 133 sampled profiles, 19 meet both conditions. The comparison concerns this finite spectral family.}
		\label{fig:app_onepole}
	\end{figure}
	
	\begin{figure}[t]
		\centering
		\includegraphics[width=0.49\textwidth]{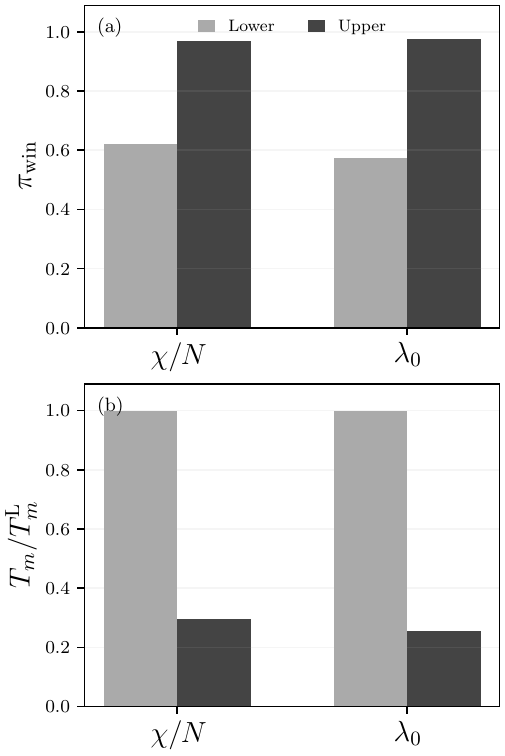}
		\caption{Comparison of the two $J_z^2$ interaction continuations at $N=12$ and matched stationary excitation fraction $\langle k\rangle/N=0.42$. (a) Stationary target-window probability $\pi_{\rm win}$ for $|k-m|\leq2$ for the two distinct cutoff-center solutions within each scheme. (b) Mean first-passage-time ratio $T_m/T_m^{\rm L}$, with $T_m^{\rm L}$ referring to the lower-center solution, under the same effective upward-rate budget and weak-rate scaling. The extensive scheme uses $\lambda_N=\chi/N$, while the fixed-pair scheme uses $\lambda_0=0.08$; each has strictly ascending transitions at this finite size. The comparison demonstrates a model-level phenomenon under both prescriptions, without extrapolating the fixed-pair model beyond its positive-gap window.}
		\label{fig:app_two_scalings}
	\end{figure}
	
	\section{Spin-sector leakage and symmetry-breaking charging}\label{app:local_robustness}
	
	For continuous charging with pure local dephasing, the state $\rho_\infty=Z_N^{-1}\sum_k a_k\Pi_k$ is stationary under both the collective and local contributions to the generator. Each $\Pi_k$ is invariant under dephasing, while the collective flux between neighboring excitation manifolds vanishes because $a_{k-1}S_+(\Omega_k)=a_kS_-(\Omega_k)$. With positive rates along every adjacent edge and independent dephasing on all spins, the stationary state is also unique. To see this, $\rho_\infty$ has full rank, and operators commuting with every local $\sigma_z^{(i)}$ are diagonal in the computational basis. Commutation with the resolved collective raising and lowering operators then equates the entries of configurations connected by a single spin flip. These connections span the full computational basis, leaving only scalar operators in the commutant. Hence the generator has a unique full-rank stationary state. The identity $\operatorname{Tr}\Pi_k=\binom Nk$ yields the excitation-manifold probabilities and maximal-spin fraction in Eq.~\eqref{eq:leakage_stationary}.
	
	The approach to this symmetry-diluted state can nevertheless be slow. In the permutation-invariant population dynamics, the smallest nonzero decay rate $\Delta_{\rm pop}$ is approximately proportional to weak local dephasing. At $N=20$ and $\gamma_\phi/\Gamma_{\rm res}=10^{-4}$, its values for the lower- and upper-center reservoirs are approximately $(4.8,7.0)\times10^{-7}$, corresponding to inverse-gap times of order $10^6$, far longer than a charging contact. The population-generator result does not determine the coherence spectrum of the full quantum Liouvillian. At zero dephasing the total-spin sectors are dynamically invariant; any positive local-dephasing rate eventually produces the unique mixed-sector stationary state. Thus the limits of long time and vanishing dephasing need not commute. The continuously charged stationary state also differs from the periodic state of a battery undergoing repeated extraction.
	
	The collective Hamiltonian and the resolved pump/loss jumps preserve total spin. In the extended model their action in a spin-$J$ sector is obtained by replacing the maximal-spin factor by $C_{J,m}^{+}=(J-m)(J+m+1)$ and $C_{J,m}^{-}=(J+m)(J-m+1)$, evaluated at the same edge frequency $\Omega_k$, with $k=m+N/2$. Independent local emission, excitation, and dephasing act through
	\begin{equation}\label{eq:local_noise_generator}
		\mathcal L_{\rm loc}\rho
		=\sum_{i=1}^N\left[\gamma_\downarrow\Diss[\sigma_-^{(i)}]
		+\gamma_\uparrow\Diss[\sigma_+^{(i)}]
		+\gamma_\phi\Diss[\sigma_z^{(i)}]\right]\rho,
	\end{equation}
	where $\Diss[L]\rho=L\rho L^\dagger-\frac12\{L^\dagger L,\rho\}$. For the symmetric Dicke state $\ket k$, its initial loss rate from the maximal-spin sector is exactly
	\begin{align}\label{eq:local_sector_leak}
		-\dot p_{\rm sym}(0)={}&\frac{\gamma_\downarrow}{N}k(k-1)
		+\frac{\gamma_\uparrow}{N}(N-k)(N-k-1)\nonumber\\
		&+\frac{4\gamma_\phi}{N}k(N-k).
	\end{align}
	The dephasing jump is $\sigma_z$, rather than $\sigma_z/2$. At the representative occupation $N=20$, $k=8$, the coefficient of $\gamma_\phi$ in Eq.~\eqref{eq:local_sector_leak} is larger than those multiplying the individual excitation and emission rates. Consequently, local dephasing can redistribute charge between total-spin sectors even while leaving the excitation number unchanged.
	
	The population reduction has $(N/2+1)^2=121$ distinct $(J,m)$ coordinates for $N=20$. The calculations cover $\gamma_\phi/\Gamma_{\rm res}\in[10^{-6},10^{-3}]$, with $\Gamma_{\rm res}=0.04\Delta\Omega$ and up to $K=1000$ charge--extract cycles. The reduced local transition rules agree with a full $N=4$ spin-space construction, and the collective reservoir rates retain the same weak-rate normalization across sectors.
	
	For the all-sector work operation the pulse is generated by $H_B+\Omega_xJ_x$, with $\Omega_x=1$ and $t_{\rm w}=1.5$ in every total-spin sector. The clean-work-matched contact durations are $\tau_{\rm L}=11765$ and $\tau_{\rm U}=2973.6$, with common clean mean work $\mu_0\simeq0.145$. Each noisy sequence begins from the clean periodic post-extraction state of this same transverse pulse. Its energy-basis TPM kernel is $D_{(J,m'),(J,m)}=|\langle J,m'|\exp[-it_{\rm w}(H_B+\Omega_xJ_x)]|J,m\rangle|^2$.
	
	At the representative \emph{equal physical dephasing rate}, the lower-center protocol delivers almost no mean work after a hundred cycles, while the upper-center protocol retains appreciable positive output. The mean work per cycle is approximately $0.003$ and $0.13$, respectively, and the fractions remaining in the maximal-spin sector are about $6\%$ and $48\%$. Over indefinitely repeated operation, both protocols approach a symmetry-diluted cyclic regime with maximal-spin weight of order $10^{-5}$ and negative mean output. The finite-use contrast is therefore governed by how rapidly accessible charge redistributes between sectors, rather than by different asymptotic stationary laws. In these comparisons, the charging durations fixed by the clean equal-work condition remain unchanged as dephasing is introduced.
	
	Finite-horizon work statistics follow from the tilted cycle kernel $T_s=D_sC_\tau$: if $f_K(s)=\bm1^\mathsf T T_s^K\bm q_0$, then $\langle W_K\rangle=f_K'(0)$ and $\Var(W_K)=f_K''(0)-[f_K'(0)]^2$. The derivatives can be propagated together with the population state. The sector-reduced expression agrees with the full $N=4$ spin calculation, including coherent evolution during the work pulse. The same-rate, equal-integrated-dose, and equal-elapsed-time comparisons use the same pump spectra and controller; local noise acts during charging, while the pulse itself is coherent.
	
\begin{figure}[t]
		\centering
		\includegraphics[width=\columnwidth]{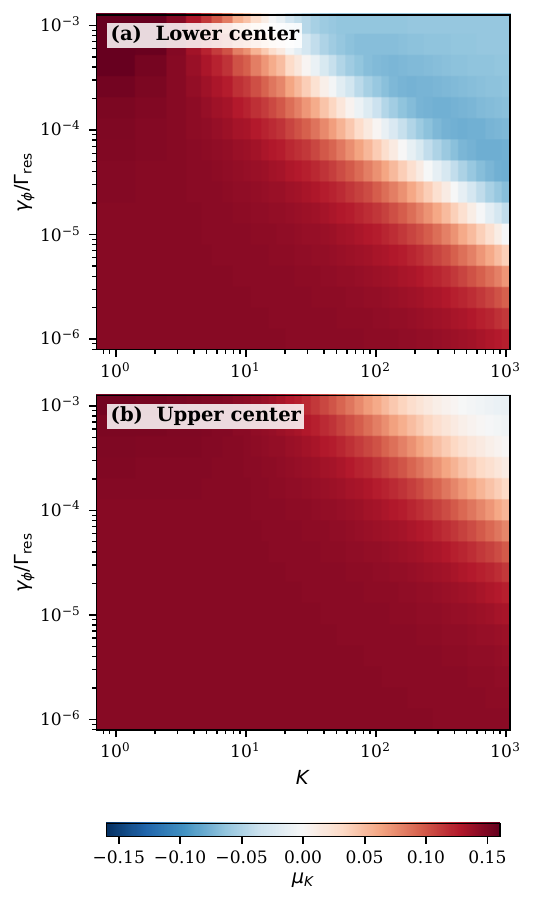}
		\caption{Accumulated mean extracted work per cycle, $\mu_K=\langle W_K\rangle/K$, versus local-dephasing strength $\gamma_\phi/\Gamma_{\rm res}$ and cycle count $K$, with (a) lower-center and (b) upper-center charging. The pulse, clean-work-matched charging times, and initial periodic states are those of Fig.~\ref{fig:local_noise_window}. The color scale distinguishes positive mean extraction from net investment. Work precision and the equal-dose comparison are displayed in the main Fig.~\ref{fig:local_noise_window}.}
		\label{fig:app_meanwork_noise_map}
	\end{figure}

	The full scan of dephasing rates and cycle number is shown in Fig.~\ref{fig:app_meanwork_noise_map}. It complements the representative finite-horizon precision comparison of Fig.~\ref{fig:local_noise_window} by locating the onset of mean work investment across the calibrated noise interval.
	
	A complementary exact-spin calculation tests local static frequency disorder acting \emph{during charging}. We use $N=6$ and add
	\begin{equation}\label{eq:disordered_charge_h}
		H_{\rm dis}=\frac12\sum_{i=1}^N\delta_i\sigma_z^{(i)},
		\qquad \sum_i\delta_i=0,
	\end{equation}
	to the original $\Omega_k$-resolved collective-jump GKSL generator. The calculation acts on the full $2^6=64$-dimensional spin Hilbert space, including coherent mixing between spin sectors. The spectral jump operators are those of the unperturbed battery: disorder modifies the Hamiltonian without reconstructing the reservoir coupling in the disordered eigenbasis. Both reservoirs are sped up by a common factor of $20$ relative to the weak-rate $N=6$ reference, leaving the unperturbed edge-rate ratios unchanged. The resulting escape-rate-to-spacing ratios are both about $0.02$.
	
	Each point uses six deterministic zero-mean detuning patterns, evolved from the empty state through a contact of duration $t=1803$. The cutoff centers are inherited from the $N=20$ reference; at $N=6$ the unperturbed mean excitations consequently differ. This isolates the sensitivity of the symmetry content to static detuning rather than comparing two chargers at equal mean occupation.
	
	\begin{figure}[t]
		\centering
		\includegraphics[width=\columnwidth]{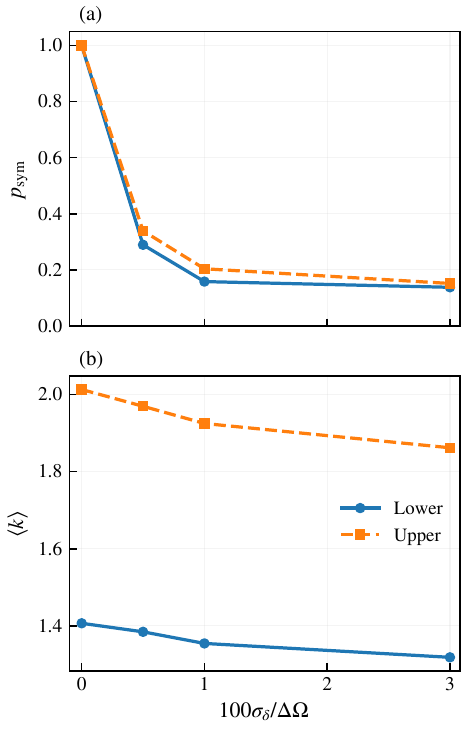}
		\caption{Static longitudinal disorder during collective charging in the full $N=6$ spin model, evaluated after a contact $t=1803$. (a) Median maximal-spin population $p_{\rm sym}$ over six zero-mean disorder realizations for each nonzero detuning strength. (b) Mean excitation $\langle k\rangle$, averaged over realizations. The horizontal axis $100\sigma_\delta/\Delta\Omega$ expresses disorder as a percentage of adjacent-transition spacing. The pump centers come from the $N=20$ reference and are not rematched at $N=6$. At a detuning of $0.5\%$ of that spacing, roughly one-third of the population remains in the maximal-spin sector while the mean excitation changes little.}
		\label{fig:app_disorder_charging}
	\end{figure}
	
	Even when the detuning is only half a percent of the adjacent-transition spacing, population leaks strongly from the maximal-spin sector while the mean excitation remains close to its disorder-free value (Fig.~\ref{fig:app_disorder_charging}). Static frequency disorder therefore changes the symmetry content of the stored charge before substantially changing the charge itself. The reduction results from coherent intersector mixing during the contact and depends on the individual detunings and evolution time; this finite-size example does not determine a disorder threshold for the twenty-spin battery.
	
	\section{Numerical controls and scope of the comparison}\label{app:valid}
	
	The stationary population ratios, first-passage recursion, and cycle-counting relations each admit independent numerical representations. Flux balance agrees with the stationary generator, first-passage times with the absorbing-generator inverse, and work cumulants with both finite-cycle moments and the covariance decomposition. Gross reservoir-energy transfers satisfy the charging-energy balance; the zero-contact exchange gives exact cancellation between successive work strokes.
	
	The equal-capacity comparison fixes $B_N$, the sum of upward edge rates, while the resolved-transition ceiling $\max_k(u_k+d_k)\leq0.04\Delta\Omega$ sets the admissible dissipative scale. Their physical content is kinetic: a complete source-energy comparison additionally depends on how the active pump is maintained. The single-Lorentzian comparison explores a bounded spectral family, with the attainable values determined by that family and the matched-work condition.

	\paragraph*{Spectral calibration and partial-secular cross terms.}
	The two cutoff profiles retain the same upward-rate capacity while the upper-center cutoff is shifted by $|\delta c|\leq0.25\Delta\Omega$ and its width by up to $20\%$. Its contact duration is retimed to the unchanged lower-center mean-work reference. Across the 33 sampled profiles, the long-time work variance remains $36$--$53\%$ smaller and the charging contact $81$--$87\%$ shorter. The negative-covariance advantage therefore persists over these finite spectral variations of the resolved-edge model.
	
	A completely positive coarse-grained extension coupling distinct adjacent ladder jumps leaves the population generator unchanged because each jump has a different final state and off-diagonal cross-jump products vanish. Its additional terms affect coherences rather than the population-based observables evaluated here. Microscopic corrections to the effective transition rates and the bath-correlation dynamics are distinct from that algebraic cross-jump effect.
	
	\bibliography{references}
\end{document}